\documentclass[twocolumn]{jpsj3}
\usepackage{txfonts}
\usepackage{amsmath,amssymb,amsfonts,float}
\newcommand{\bolS}{\text{\bf S}}

\newcommand{\bolk}{\mathbf{k}}

\newcommand{\bolQ}{\mathbf{Q}}
\newcommand{\boln}{\mathbf{n}}

\newcommand{\boll}{\mathbf{l}}

\newcommand{\boli}{\mathbf{i}}
\newcommand{\bolj}{\mathbf{j}}
\newcommand{\bolv}{\mathbf{v}}

\newcommand{\bra}[1]{\langle #1 |}  
\newcommand{\ket}[1]{| #1 \rangle}  
\newcommand{\VEV}[1]{\langle #1 \rangle}  

\title{Reflection and refraction process of spinwave in a ferromagnet/frustrated ferromagnet junction system}
\author{Yuta \name{Sasaki} and Hiroaki T. \name{Ueda}}
\inst{Department of Physics, Tokyo Metropolitan University, Hachioji, Tokyo 192-0397, Japan} 

\abst{Frustration introduces a nontrivial dispersion relation 
of spinwave even in a ferromagnetic phase in a spin system. 
We study the reflection and refraction process of spinwaves in the 
ferromagnet/frustrated ferromagnet junction system
by using the Holstein-Primakoff spinwave expansion and 
taking the large-$S$ limit. 
We discuss the relation between the incident angle 
and the refraction angle of spinwave, namely, the Snell's law of spinwave. 
As concrete examples of frustrated ferromagnets, 
we study the fully polarized ferromagnet phases 
in $J_1$-$J_2$ chains and the $J_1$-$J_2$ models on the square lattice. 
The interesting refraction processes, such as the splitting of the 
incident spinwave and the negative refraction, are discussed. 
We also study the transmittance and reflectance in these concrete models. }

\kword{spinwave, magnonics, junction system, frustration, negative refraction}

\begin{document}
\maketitle

\section{Introduction}
Spinwave is a collective excitation of spins in magnets. 
Since spinwave can carry information on the nano/microscale, 
recently, potential application of spinwave to new devices 
has attracted much attention to researchers.\cite{Magnonics} 
In many circumstances, spinwave behaves like a sound or light wave; 
reflection and refraction process of spinwave can be considered. 

Understanding of the reflection and refraction process is important 
to control spinwave. 
In the junction system of usual ferromagnets 
in which spinwave has a `trivial'\cite{foot_0} dispersion relation, 
the reflection and refraction process is theoretically
understood by using geometrical-optics approximation
\cite{Goedsche,Gorobets,Reshetnyak,Reshetnyak2} 
or Landau-Lifshitz-Gilbert equation\cite{negative_spin}. 
One of the important results of these studies is 
that the direction of the propagating spinwave depends on 
the strength of applied magnetic field. 
The experimental control of propagation of spinwave is indeed 
realized by tuning the internal demagnetizing fields in a permalloy waveguid
\cite{Demidov}. 
If the dispersion relation of spinwave is nontrivial, 
an exotic refraction process is expected. 
In the case of light, for example, the negative refraction occurs due to the nontrivial dispersion relation.\cite{Veselago,meta} 
Recently, it is theoretically proposed that 
the anisotropic nature of the dispersion relation of dipole-exchange spinwaves
can result in the negative refraction\cite{negative_spin}. 

Recent experimental advance has been exhibiting 
a wide variety of frustrated magnets. 
Magnetic frustration provides us with an exotic magnetic behavior.\cite{Diep} 
Even in a ferromagnet, in which all spins align, the frustration can induce an unusual dispersion relation of spinwaves. 
Hence, it will be natural to expect interesting refraction processes of spinwaves in frustrated ferromagnets. In this paper, we study the reflection and refraction process of spinwave in ferromagnet/frustrated ferromagnet junction system. 

As a concrete example of frustrated ferromagnets, 
we consider the spin systems 
which are already realized experimentally. 
For example, a frustrated ferromagnet can be prepared by 
applying high magnetic field on every frustrated magnet. 
One-dimensional (1D) $J_1$-$J_2$ Heisenberg chain with a nearest-neighbor exchange coupling $J_1$ and next-nearest-neighbor coupling 
$J_2$ is one of the famous frustrated magnets\cite{Nagamiya,Hase,Chubukov,Hikihara1,Hikihara2}. 
In the fully polarized phase under high magnetic field, 
the dispersion of spinwave has two minima at\cite{Nagamiya,Kuzian,Ueda_Totsuka,Kolezhuk} the wavevector $Q$s of $\cos Q=-J_1/4J_2$ for $|J_1|/J_2<4$ and $J_2>0$. 
There are many compounds consisting of the $J_1$-$J_2$ chains 
(see Table.~I in Ref.~\citen{Hase}). 
The saturated ferromagnetic phases under high magnetic field are experimentally realized in \cite{Hase}Rb$_2$Cu$_2$Mo$_3$O$_{12}$, \cite{Brown,Hagiwara,Maeshima}(N$_2$H$_5$)CuCl$_3$ and \cite{LiCuVO4_1,LiCuVO4_2,LiCuVO4_3}LiCuVO$_4$. 

As another example of the frustrated magnets, the various compounds of 
the square-lattice $J_1$-$J_2$ model, e.g., BaCdVO(PO$_4$)$_2$, are reported\cite{J1J2compound_1,J1J2compound_2}. Without external field, the collinear antiferromagnet phase appears in BaCdVO(PO$_4$)$_2$; the saturation field is about 4$\sim$6T. 
In the saturated ferromagnetic phase of this compound, theoretically, the dispersion relation is expected to be nontrivial; it has two minima at $\bolQ=(0,\pi)$ and $\bolQ=(\pi,0)$.\cite{J1J2compound_1,J1J2compound_2,Thalmeier}

The organization of the present paper is as follows. 
In Sec.~\ref{Sec:disp}, 
we briefly review the dispersion relations of spinwaves 
in (frustrated) ferromagnets by using the Holstein-Primakoff 
spiwave expansion. 
In Sec.~\ref{Sec:Snell}, 
we generally discuss refraction angles of 
transmission spinwaves in the ferromagnets junction systems 
satisfying the given conditions. 
Then, we apply this discussion 
to the several junction systems including the frustrated 
$J_1$-$J_2$ chains or the $J_1$-$J_2$ model on the square lattice. 
We shall explicitly see the splitting of the incident spinwave 
and the negative refraction in these models. 
In Sec.~\ref{reftra}, we study 
the reflection- and the transmission rates (reflectance and transmittance)
in part of the junction systems considered in Sec.~\ref{Sec:Snell} 
by solving the simple Schr\"{o}edinger equations within the large $S$ limit. 
Our approach naturally treats the lattice structure of ferromagnets 
without the coarse graining.

\section{Dispersion Relation of Frustrated Ferromagnets}
\label{Sec:disp}
As a brief review, let us discuss the dispersion relation of spinwaves 
in the fully-polarized ferromagnetic phase of (frustrated) spin systems. 
For simplicity, we consider the lattice systems with one magnetic ion per unit cell, and assume the rotational symmetry around the $z$ direction in spin space. 
This assumption leads to 
the conservation law of the total angular momentum along the $z$ direction, 
namely, $\sum_\boli \VEV{S^z_\boli}$.
We study the Hamiltonian with the generic exchange interactions 
$J_{\boli\bolj}=J_{\boli-\bolj}=J_{\bolj-\boli}=J_{\boll}$, 
the on-site anisotropic interaction term $K$, and the external magnetic field $H$:
\begin{equation}
H=\sum_{\VEV{\boli,\bolj}}J_{\boli \bolj}{\bf S}_{\boli}\cdot {\bf S}_{\bolj}
+\sum_{i}(-K(S_{\boli}^{z})^2+{\rm H}S_{\boli}^z)\ ,
\label{HeisenbergH}
\end{equation}
where we use the coordinate $(x,y,z)$ in spin space, and $\boli=(a_i,b_i,c_i)$ in lattice space. 
Let us rewrite the spin operator by using the Holstein-Primakoff transformation on the fully-polarized ferromagnetic phase:
\begin{equation}
\begin{split}
S^z_\boli&=-S+\alpha^\dagger_\boli \alpha_\boli\ ,\\
S_\boli^+ &=\sqrt{2S} \alpha_\boli^\dagger \sqrt{1-\frac{ \alpha^\dagger_\boli \alpha_\boli}{2S}}\approx \sqrt{2S} \alpha_\boli^\dagger \ ,\\
S_\boli^- &=\sqrt{2S}\sqrt{1-\frac{ \alpha^\dagger_\boli \alpha_\boli}{2S}} 
\alpha_\boli\approx \sqrt{2S} \alpha_\boli\ .
\end{split}
\label{Holst}
\end{equation}
By using the $1/S$ expansion, we approximately obtain the free bosonic Hamiltonian in the leading order in $1/S$:
\begin{equation}
H=\sum_\bolk (\omega(\bolk)-\mu)\alpha^\dagger_\bolk \alpha_\bolk\ ,
\end{equation}
where
\begin{equation}
\begin{split}
\epsilon(\bolk)&=\frac{1}{2}\sum_\boll J_\boll \cos\bolk\cdot \boll\ ,
\ \ \omega(\bolk)=2S(\epsilon(\bolk)-\epsilon_{\text {min}})\ ,\\ 
\mu&=2S(\epsilon(0)-\epsilon_\text {min})-({\rm H}+2SK)\ ,
\end{split}
\label{dispersion}
\end{equation}
and $\epsilon_{\text {min}}$ is the minimum of $\epsilon(\bolk)$. 
The conservation law of the angular momentum along the $z$ direction 
assures the conservation of the total magnon number. 
For any exchange interactions, 
a sufficiently large ${\rm H}$ or $K$ can induce the gap ($\mu\leq0$) in the magnon dispersion. 
If $\mu \leq0$, the ferromagnetic phase is stable. 
Throughout this paper, we focus on this free bosonic Hamiltonian 
by assuming the large $S$ limit. 

We have seen that a dispersion relation takes various forms due to frustration. 
This dispersion relation can lead to the nontrivial group velocity not parallel to the phase velocity, where the group velocity is given by
\begin{equation}
\bolv_g(\bolk)=\nabla_\bolk \omega(\bolk)\ .
\end{equation}
The group velocity has an important physical meaning: 
it carry the angular momentum as reviewed in Appendix. 
Not the phase- but the group velocity determines 
the traveling direction of spinwave. 
Next, let us study the various dispersion relations in the concrete models.

\subsection{simple case}
\label{Sec:simple}
\begin{figure}[ht]
\begin{center}
\includegraphics[scale=0.25]{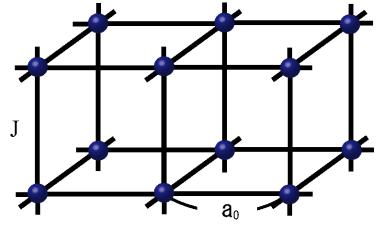}
\caption{(Color online)
The cubic lattice 
with the nearest neighbor Heisenberg exchange coupling $J$. 
The dots represent spins. The lattice constant $a_0=1$ is assumed.
\label{Fig:cubic}}
\end{center}
\end{figure}

\begin{figure}[ht]
\begin{center}
\includegraphics[scale=0.7]{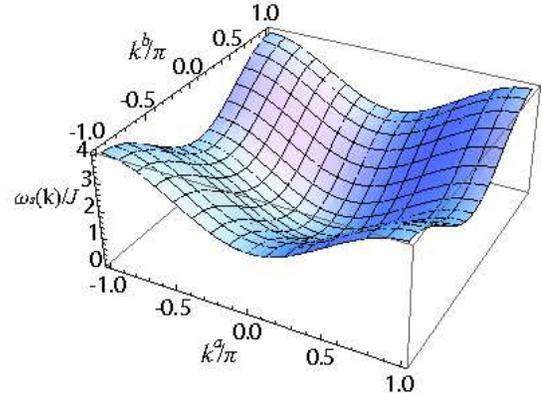}
\caption{(Color online)
The dispersion relation $\omega_s(\bolk)$ of the trivial ferromagnet on the cubic lattice for $J<0$. $k^c=0$ is assumed.
\label{Fig:dispFM}}
\end{center}
\end{figure}

First, we consider the simple Heisenberg model on the cubic lattice 
with the nearest neighbor ferromagnetic coupling $J<0$ as shown in Fig.~\ref{Fig:cubic}. 
$\epsilon(\bolk)$ is given by
\begin{equation}
\begin{split}
\epsilon_s(\bolk)&=J(\cos k^a + \cos k^b + \cos k^c)\ ,\\
\epsilon_{s\text{min}}&=\epsilon_s(\bolk_0=(0,0,0))=3J\ .
\end{split}
\end{equation}
This leads to the dispersion relation $\omega_s(\bolk)$ shown in Fig.~\ref{Fig:dispFM}. 
In the long wavelength limit $|\bolk|\rightarrow 0$, 
this dispersion relation becomes isotropic and is given by 
\begin{equation}
\omega_s(\bolk)\approx\frac{\bolk^2}{2m_s}\ ,\ \ m_s=-\frac{1}{J}\ .
\end{equation}
In the following discussion, we use this model as the concrete example of `usual' ferromagnets. 

\subsection{$J_1$-$J_2$ chain}
\label{Sec:J1J2}
\begin{figure}[ht]
\begin{center}
\includegraphics[scale=0.25]{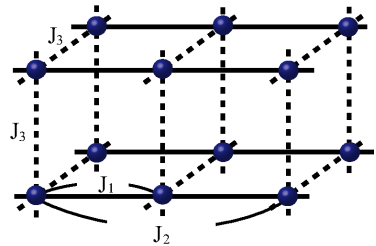}
\caption{(Color online)
$J_1$-$J_2$ chains with the interchain coupling $J_3$ on the cubic lattice. The dots represent spins. 
\label{Fig:J1J2J3}}
\end{center}
\end{figure}

\begin{figure}[ht]
\begin{center}
\includegraphics[scale=0.6]{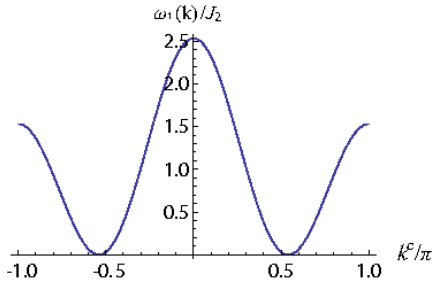}
\caption{(Color online)
The dispersion relation $\omega_1(\bolk)$ of the 1D $J_1$-$J_2$ chain
for $J_1/J_2=0.5$ and $J_2>0$. $J_3=0$ or $k^a=k^b=0$ is assumed.
\label{Fig:disp_chain}}
\end{center}
\end{figure}

\begin{figure}[ht]
\begin{center}
\includegraphics[scale=0.6]{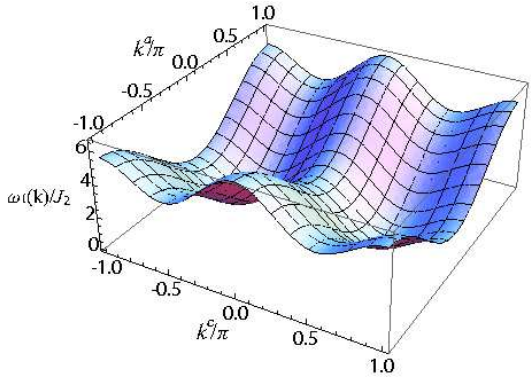}
\caption{(Color online)
The dispersion relation $\omega_1(\bolk)$ of the 1D $J_1$-$J_2$ chains on the cubic lattice for $J_1/J_2=0.5$, $J_2>0$ and $J_3/J_2=-2$. $k^c=0$ is assumed.
\label{Fig:disp_chain2}}
\end{center}
\end{figure}

Next, let us discuss the dispersion relation of 
the 1D $J_1$-$J_2$ chains with the ferromagnetic interchain coupling $J_3<0$ 
on the cubic lattice\cite{Kuzian,Ueda_Totsuka,Kolezhuk,Nishimoto} as illustrated in Fig.~\ref{Fig:J1J2J3}: 
\begin{equation}
\begin{split}
\epsilon_1(\bolk) & =J_1 \cos k^c+J_2\cos 2k^c+J_3(\cos k^a+\cos k^b)\\
& =2J_2 (\cos k^c+\frac{J_1}{4J_2})^2-\frac{J_1^2}{8J_2}-J_2
+J_3(\cos k^a+\cos k^b)\ ,
\end{split}
\end{equation}
where the $J_1$-$J_2$ chains are assumed to lie parallel to the $c$ axis. 
For $|J_1|/J_2 \leq 4$, $J_2>0$ and $J_3<0$, the dispersion relation 
has two minima at $\bolQ=(0,0,\pm Q^c)$ where
\begin{equation}
\cos Q^c =-\frac{J_1}{4J_2}\ ,\ \ 
\epsilon_\text{min}=-\frac{J_1^2}{8J_2}-J_2+2J_3\ .
\end{equation}
The dispersion relation $\omega_1(\bolk)$ is graphically shown 
in Figs.~\ref{Fig:disp_chain},\ref{Fig:disp_chain2}.

\subsection{$J_1$-$J_2$ model on the square lattice}
\begin{figure}[ht]
\begin{center}
\includegraphics[scale=0.24]{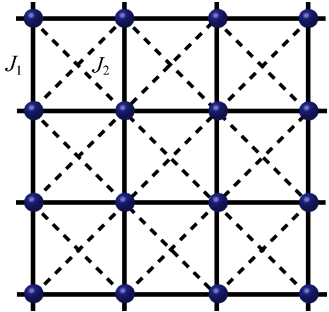}
\caption{(Color online) 
Two dimensional square lattice. 
The dots represent spins connected by 
the nearest neighbor Heisenberg exchange coupling $J_1$ 
and the next nearest neighbor coupling $J_2$. 
\label{Fig:square}}
\end{center}
\end{figure}

\begin{figure}[ht]
\begin{center}
\includegraphics[scale=0.6]{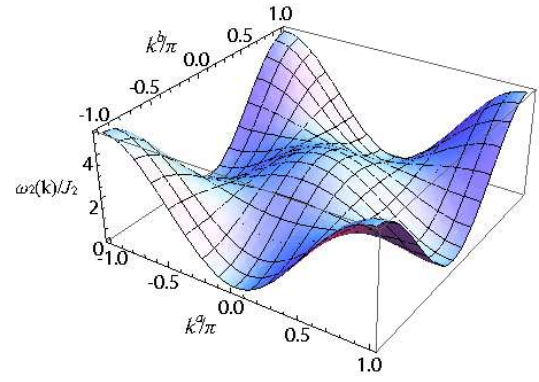}
\caption{(Color online)
The dispersion relation $\omega_2(\bolk)$ for $J_1/J_2=-0.5$ and 
$J_2>0$
in the $J_1$-$J_2$ model on the two-dimensional square lattice. 
\label{Fig:disp_square}}
\end{center}
\end{figure}

The fully saturated phase in the $J_1$-$J_2$ model on the square lattice 
(see Fig.~\ref{Fig:square}) 
also has the nontrivial dispersion relation:
\begin{equation}
\begin{split}
\epsilon_2(\bolk) & =J_1 (\cos k^a+\cos k^b)+J_2(\cos (k^a+k^b)+\cos (k^a-k^b)).
\end{split}
\label{J1J2disp}
\end{equation}
For $-2<J_1/J_2<2$ and $J_2>0$, 
the dispersion relation $\omega_2(\bolk)$ has two minima at 
$\bolQ^{(2)}_1=(0,\pi)$ and $\bolQ^{(2)}_2=(\pi,0)$
 as shown in Fig.~\ref{Fig:disp_square}. 


\section{The Snell's Law in the Ferromagnets Junction System}
\label{Sec:Snell}
\begin{figure}[ht]
\begin{center}
\includegraphics[scale=0.5]{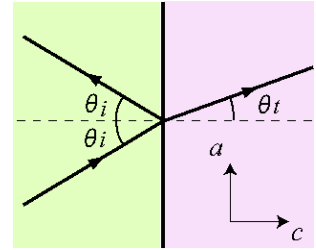}
\caption{(Color online)
Schematic figure of refraction and reflection process of 
spinwave in the usual ferromagnets junction system 
in the case of $k_b=0$. 
The travelling direction of spinwave is determined by the group velocity.
\label{Fig:FM11}}
\end{center}
\end{figure}

In this section, we study the relation of the angles of 
incident-, reflected- and transmission-
spinwaves passing through a boundary (the Snell's law). 
Let us consider a ferromagnet/ferromagnet 
junction system whose boundary plane is flat and 
is perpendicular to $c$ direction. 
On the boundary the proximity effect may induce 
a spin-exchange coupling which magnetically relates two ferromagnets. 
This proximity effect leads to transmission of incident spinwave. 
The schematic figure of the reflection and refraction process 
in the usual ferromagnets junction system is shown in Fig.~\ref{Fig:FM11}.

In the case of light, the Snell's law is 
understood by the conservation law of the energy and 
that of the momentum parallel to the surface. 
The same context can be applied to determines the Snell's law of spinwave 
in the ferromagnets junction system
if the following conditions are satisfied:
(i) The exchange interactions in each ferromagnet and on the boundary have a rotational symmetry around the $z$ direction in spin space. 
(ii) Each ferromagnet has a translational symmetry (far away from the boundary).
(iii) Both ferromagnets have the same lattice vectors in the $a$-$b$ plane, 
and 
the translational symmetry in the $a$-$b$ plane exists 
in the whole junction system. 
\cite{footnote}
The assumption (i) leads to the conservation law of a total number of magnons, 
and  (ii),(iii) lead to the conservation law of a total momentum of the $a$ and $b$ direction. 
Of course, the momentum conservation law of the $c$ direction does not hold by the boundary effect. 

\begin{figure}[ht]
\begin{center}
\includegraphics[scale=0.5]{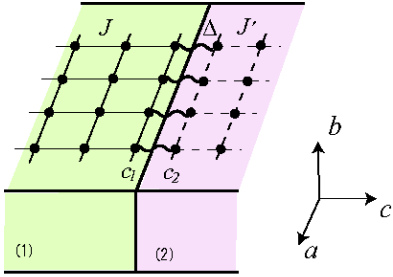}
\caption{(Color online) 
Junction system of usual ferromagnets on the cubic lattice. The dots represent spins. The exchange coupling $\Delta$ due to the quantum proximity effect 
is considered on the boundary.
\label{Fig:FMFM}}
\end{center}
\end{figure}

Since the above discussion may be formal, let us consider the simple junction system consists of ferromagnets with the nearest neighbor interaction $J$ ($J^\prime$) on the cubic lattice (see Sec.~\ref{Sec:simple}) with the same lattice constant. 
This junction system is schematically illustrated in Fig.~\ref{Fig:FMFM}. 
It may be appropriate to consider the exchange interaction on the boundary as
\begin{equation}
\Delta\sum_{a,b}\bolS_{\boli=(a,b,c_1)} \cdot \bolS_{\bolj=(a,b,c_2)} \ ,
\label{SS:boundary}
\end{equation}
where $\Delta$ is the strength of the exchange interaction on the boundary, 
and $c_1$ and $c_2$ is the position of the boundary in the c direction. 
This boundary condition satisfies the conditions (i), (iii).\cite{footnote_boundary} 

Let us consider the case that the incident spinwave in the medium (1) 
has the momentum $\bolk_i$. 
The incident spinwave arriving at a boundary 
is divided into the transmission wave of $\bolk_t$ 
and the reflected wave of $\bolk_r$. 

The assumption (i) leads to the energy conservation 
\begin{equation}
\Omega=\omega_{(1)}(\bolk_i)-\mu_{(1)}=\omega_{(1)}(\bolk_r)-\mu_{(1)}
=\omega_{(2)}(\bolk_t)-\mu_{(2)}\ . 
\label{Econserv}
\end{equation}
As discussed in Appendix, 
the group velocity corresponds to the current of magnons. 
Hence, it is appropriate 
in our setting to assume that the group velocity of 
the transmission (reflection) wave has the same (opposite) direction along the c axis as that of the incident wave. Namely,
\begin{equation}
v_{g}^{(1)c}(\bolk_i)/
|v_{g}^{(1)c}(\bolk_i)|
=-v_{g}^{(1)c}(\bolk_r)/|v_{g}^{(1)c}(\bolk_r)|
=v_{g}^{(2)c}(\bolk_r)/|v_{g}^{(2)c}(\bolk_r)|\ .
\label{groupD}
\end{equation}

In addition, the momentum conservation law parallel to the boundary 
leads to 
\begin{equation}
k_i^a=k_r^a=k_t^a\ ,\ \ 
k_i^b=k_r^b=k_t^b\ .
\label{abcons}
\end{equation} 
These equations determine $\bolk_r$ and $\bolk_t$, 
which lead to the Snell's law. 
In general, the transmission- and reflected spinwaves determined by eqs.~(\ref{Econserv}), (\ref{groupD}), (\ref{abcons}) are not necessarily single valued. 

Finally, let us discuss the general relation between the reflectance and transmittance if the conditions (i) (ii) (ii) are satisfied. 
By labeling spinwaves as $\bolk^{(1)},\bolk^{(2)},\bolk^{(3)} \cdots$, this reflection and refraction process can be written as the ket
\begin{equation}
(A\alpha_{\bolk_i}^{(1)\dagger }+
\sum_jB_j\alpha_{\bolk^{(j)}_{r}}^{(1)\dagger }+
\sum_jC_j\alpha_{\bolk^{(j)}_{t}}^{(2)\dagger })\ket{0}\ ,
\end{equation}
where $\ket{0}$ is vacuum state and
$a^{(1,2)}(\bolk)$ approaches the free boson in the media (1,2) with the momentum $\bolk$ sufficiently away from the boundary. 
$|v_{g}^{(1)c}(\bolk^{(j)}_{r})/v_{g}^{(1)c}(\bolk_i)||B_j/A|^2$ ($|v_{g}^{(2)c}(\bolk^{(j)}_{t})/v_{g}^{(1)c}(\bolk_i)||C_j/A|^2$) is reflectance (transmittance).
By considering the cuboid discussed in Appendix, we obtain 
the magnon number conservation law: 
\begin{equation}
|A|^2v_{g}^{(1)c}(\bolk_i)+\sum_j|B_j|^2v_{g}^{(1)c}(\bolk^{(j)}_{r})
=\sum_j|C_j|^2 v_{g}^{(2)c}(\bolk^{(j)}_{t})\ .
\label{conserv_s}
\end{equation}
The concrete expression of $B_j$ and $C_j$ will be discussed 
in Sec.~\ref{reftra} by explicitly considering the boundary condition.

Since we have discussed the procedure to find the Snell's law,
let us study the refraction processes in the various ferromagnets junction systems. 

\subsection{usual case}
\label{Sec:usual}
First, let us further study the Snell's law 
of the junction system of the usual ferromagnets on the cubic lattice with 
the nearest neighbor exchange coupling $J<0$ ($J^\prime<0$) 
as illustrated in Fig.~\ref{Fig:FMFM}. 
The dispersion relation on each ferromagnet is respectively given by
\begin{subequations}
\begin{align}
\omega_{s1}(\bolk)&=2S(J(\cos k^a + \cos k^b + \cos k^c)-3J)\ ,\\
\omega_{s2}(\bolk)&=2S(J^\prime(\cos k^a + \cos k^b + \cos k^c)-3J^\prime), \
\end{align}
\end{subequations}
where $\mu_1<0$ and $\mu_2<0$. 
$\mu_{1,2}$ are given in eq.~(\ref{dispersion}). 
In these usual ferromagnets, the group velocity ${\bf v}_g$, 
which carry the angular momentum, 
is given by
\begin{subequations}
\begin{align}
{\bf v}_g^{(s1)}(\bolk)&=2S\nabla_\bolk \epsilon(\bolk)
=-2SJ(\sin k^a, \sin k^b, \sin k^c)\ ,\\
{\bf v}_g^{(s2)}(\bolk)&
=-2SJ^\prime(\sin k^a, \sin k^b, \sin k^c)\ .
\end{align}
\end{subequations}
The sign of the group velocity $v_g^{(s1)a,b,c}$ is always the same 
as the sign of $k^{a,b,c}$ for $-\pi<k^{a,b,c}<\pi$. 
If we consider the incident wave of wavevector $\bolk_{i}$, 
eqs.~(\ref{Econserv}), (\ref{groupD}), (\ref{abcons}) 
uniquely determine the reflected wave of $\bolk_{r}$ and 
the transmission wave of $\bolk_t$. 
We obtain 
\begin{equation}
\begin{split}
\cos k_{t}^c =& \frac{1}{J^\prime}(J\cos k^c_i + (J-J^\prime)(\cos k^a_i+\cos k^b_i)-3(J-J^\prime)\\
&+\frac{\mu_2-\mu_1}{2S})\ ,
\label{nom_kt}
\end{split}
\end{equation}
where $k_{t}^c/|k_{t}^c|=k_{i}^c/|k_{i}^c|$ for $-\pi <k_{t}^c<\pi$.
This implies that, the larger the chemical potential $\mu_2$ is, the slower 
the group velocity of the transmission wave becomes. 
Since $\mu_{1,2}$ depends on the external magnetic field ${\rm H_{1,2}}$ in each compound , 
we explicitly see that the propagation of the spinwave can be controlled 
by the (local) external magnetic field as discussed in Refs.~\citen{Goedsche,Gorobets,Reshetnyak,Reshetnyak2}.  

If we consider the long-wavelength limit $|\bolk|\rightarrow 0$, 
the Snell's law in this system becomes simple. 
The group velocity is proportional to the phase velocity 
in the long wavelength limit $\bolk\rightarrow 0$. 
Up to the order of $\bolk^2$, 
the refractive index is given by
\begin{equation}
n=\frac{\sin\theta_i}{\sin\theta_t}=\frac{|\bolk_t|}{|\bolk_c|}=
\sqrt{\frac{\Omega-J^\prime+\mu_2}{\Omega-J+\mu_1}}\ ,
\label{SimpleLong}
\end{equation}
where $\theta_i$ ($\theta_t$) is the angle of incidence (refraction) 
as shown in Fig.~\ref{Fig:FM11}, 
and $\Omega$ is given in eq.~(\ref{Econserv}). 
Eq.~(\ref{SimpleLong}) is exactly the same as the result obtained in
Ref.~\citen{Gorobets}.

\subsection{$J_1$-$J_2$ chains parallel to the boundary}
\label{Sec:J1J2p1}
\begin{figure}[ht]
\begin{center}
\includegraphics[scale=0.6]{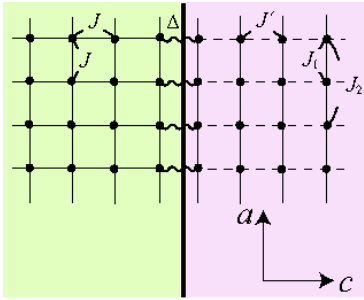}
\caption{(Color online) Junction system of the usual 
ferromagnet (left, green) /ferromagnet consisting of $J_1$-$J_2$ chains 
which lie along the $a$ axis (right, red). The exchange coupling $\Delta$ due to the quantum proximity effect is considered on the boundary.
\label{Fig:FMJ1J2a}}
\end{center}
\end{figure}

Next, let us consider the reflection and refraction process of spinwaves
in the usual ferromagnet/frustrated ferromagnet junction system. 
In this subsection, as a concrete example of frustrated ferromagnets, 
we consider the $J_1$-$J_2$ chains on the cubic lattice 
discussed in Sec.~\ref{Sec:J1J2}. 
We assume that the $J_1$-$J_2$ chains are parallel to the boundary plane 
as shown in Fig.~\ref{Fig:FMJ1J2a}. 

\begin{figure}[ht]
\begin{center}
\includegraphics[scale=0.5]{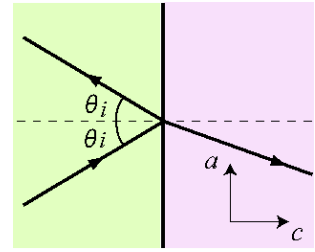}
\caption{(Color online) Schematic figure of refraction and reflection process of 
spinwave in the case of the negative refraction index and $\bolk_c=0$. 
The travelling direction of spinwave is determined by the group velocity.
\label{Fig:FMJ1J2d}}
\end{center}
\end{figure}

The dispersion relation and the group velocity 
in the $J_1$-$J_2$ chains on the cubic lattice 
are given by
\begin{equation}
\begin{split}
\omega_{11}(\bolk)=&2S(J_1\cos k^a +J_2\cos 2 k^a\\
&+ J^\prime(\cos k^b + \cos k^c -2)-\frac{J_1^2}{8J_2}-J_2)\ ,\\
{\bf v}_g^{(11)}(\bolk)=&-2S(J_1\sin k^a+2J_2\sin 2 k^a, J^\prime\sin k^b, J^\prime\sin k^c)\ .
\end{split}
\end{equation}
If we consider the incident wave of wavevector $\bolk_i$ 
in the usual ferromagnet, 
the transmission wave of $\bolk_t$ is uniquely given 
by eqs.~(\ref{Econserv}), (\ref{groupD}), (\ref{abcons}):
\begin{equation}
\begin{split}
\cos k_{t}^{c} &= \frac{1}{J^\prime}\bigl(\frac{\Omega+\mu_2}{2S}
-J_1\cos k^a_i-J_2\cos 2 k^a_i \\
&-J^\prime(\cos k^b_i-1)-\frac{J_1^2}{8J_2}-J_2\bigr)\ ,
\end{split}
\label{ktcJ1J21}
\end{equation}
where $k_{t}^c/|k_{t}^c|=k_{i}^c/|k_{i}^c|$ for $-\pi <k_{t}^c<\pi$. 
As discussed in Sec~\ref{Sec:J1J2}, 
if $-4<J_1/|J_2|<4$ and $J_2>0$, 
the equation $J_1\cos k^a_i+J_2\cos 2 k^a_i$ has two minima at 
$\cos k^a_i=-\frac{J_1}{4J_2}$. 
If $\cos k_i^a > -\frac{J_1}{4J_2}$, 
the $a$ component of the group velocities of the incident spinwave 
is of the opposite sign to that of the transmission spinwave: 
the refractive index becomes negative\cite{Murakami} as illustrated in Fig.~\ref{Fig:FMJ1J2d}.

\subsection{$J_1$-$J_2$ chains perpendicular to the boundary}
\label{Sec:J1J2p2}
\begin{figure}[ht]
\begin{center}
\includegraphics[scale=0.6]{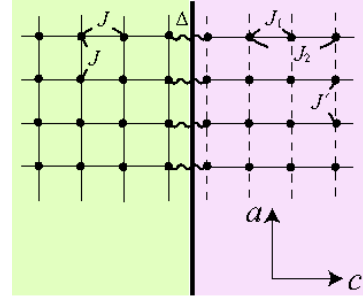}
\caption{(Color online) Junction system of 
the usual ferromagnet (left, green) /ferromagnet consisting of the $J_1$-$J_2$ 
chains which lie along the $c$ axis (right, red). 
The exchange coupling $\Delta$ due to the quantum proximity effect 
is considered on the boundary.
\label{Fig:FMJ1J2b}}
\end{center}
\end{figure}

\begin{figure}[ht]
\begin{center}
\includegraphics[scale=0.5]{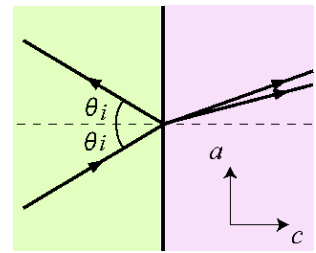}
\caption{(Color online) 
Schematic figure of refraction and reflection process of 
spinwave in the case of $\bolk_b=0$. 
The incident spinwave can split into two transmission waves. 
The travelling direction of spinwave is determined by the group velocity.
\label{Fig:FMJ1J2c}}
\end{center}
\end{figure}

In this subsection, we study the case that the frustrated ferromagnet 
consists of $J_1$-$J_2$ chains perpendicular to the boundary plane as shown in Fig.~\ref{Fig:FMJ1J2b}. 
The dispersion relation in the $J_1$-$J_2$ chains is given by
\begin{equation}
\begin{split}
\omega_{12}(\bolk)&=2S(J_1\cos k^c +J_2\cos 2 k^c\\
&+J^\prime(\cos k^a + \cos k^b-2)+\frac{J_1^2}{8J_2}+J_2)\ .
\end{split}
\end{equation}
If the spinwave of $\bolk_i$ 
in the usual ferromagnet is injected into the $J_1$-$J_2$ chains, 
the transmission waves of $\bolk_t^{(\pm)}$ are given 
by eqs.~(\ref{Econserv}), (\ref{groupD}), (\ref{abcons}):
\begin{equation}
\cos k_{t}^{(\pm)c} = \pm\sqrt{\frac{\Omega-2SJ^\prime(\cos k^a+\cos k^b-2)+\mu_2}{4SJ_2}}-\frac{J_1}{4J_2}\ ,
\label{ktcJ1J22}
\end{equation}
where $-\frac{J_1 \sin k_{t}^{(\pm)c}+2J_2 \sin 2k_{t}^{(\pm)c}}{|J_1 \sin k_{t}^{(\pm)c}+2J_2 \sin 2k_{t}^{(\pm)c}|}=\frac{k_i^c}{|k_i^c|}$. 
The appearance of $\pm$ in the right-hand side of (\ref{ktcJ1J22}) 
is because the dispersion relation has two minima in the direction of c axis.
Hence, when both $\bolk^{\pm}$ are permitted, two species of transmission 
spinwave appear as shown in Fig.~\ref{Fig:FMJ1J2c}.

\subsection{$J_1$-$J_2$ model on the square lattice}
\begin{figure}[ht]
\begin{center}
\includegraphics[scale=0.6]{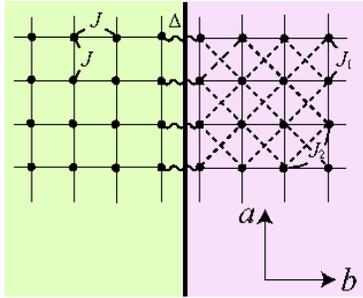}
\caption{(Color online) Junction system of 
the usual ferromagnet (left, green) /ferromagnet consisting of 
the $J_1$-$J_2$ model on the square lattice (right, red). 
The exchange coupling $\Delta$ due to the quantum proximity effect 
is considered on the boundary.
\label{Fig:FMJ1J2e}}
\end{center}
\end{figure}

Finally, let us briefly consider the fully polarized phase in the 
$J_1$-$J_2$ model on the square lattice as an example of frustrated ferromagnets. We study the 2-dimensional junction system in the $a$-$b$ plane shown in Fig.~\ref{Fig:FMJ1J2e}. 
The dispersion relation in the $J_1$-$J_2$ model is given by eq.~(\ref{J1J2disp}).
In the case that $-2 < J_1/J_2 < 2$ and $J_2>0$, by injecting the spinwave of $\bolk_i$ from the usual ferromagnet, 
the transmission spinwave of $\bolk_t$ is given by
\begin{equation}
\begin{split}
k_t^a&=k_i^a\ , \\
\cos k_t^b&=\frac{1}{J_1+2J_2\cos k_i^a}
(\frac{\Omega+\mu_2}{2S}-J_1\cos k_i^a-2J_2)\ ,
\end{split}
\end{equation}
where $-\frac{J_1\sin k^b_t+2J_2\cos k^a_t\sin k^b_t}{|J_1\sin k^b_t+
2J_2\cos k^a_t\sin k^b_t|}=-\frac{\sin k_i^b}{|\sin k_i^b|}$. 
The negative refraction can be realized as well as the junction system 
discussed in Sec.~\ref{Sec:J1J2p1} as illustrated in Fig.~\ref{Fig:FMJ1J2d}.

\section{Reflectance and Transmittance}
\label{reftra}
We have studied the relation 
between the refraction angles of transmission spinwaves 
and the angle of incident spinwave, 
namely, the Snell's law of spinwaves in ferromagnets junction system. 
In this section, let us discuss 
the reflectance and the transmittance of spinwaves. 
Although in this section we focus 
on the concrete models discussed 
in Secs.~\ref{Sec:usual}, \ref{Sec:J1J2p1}, \ref{Sec:J1J2p2}, 
our approach will be easily applied to 
other frustrated ferromagnets junction systems 
satisfying the conditions (i), (ii), (iii) 
discussed in Sec.~\ref{Sec:Snell}.

\subsection{usual case}
\label{ref_trans}
\begin{figure}[ht]
\begin{center}
\includegraphics[scale=0.5]{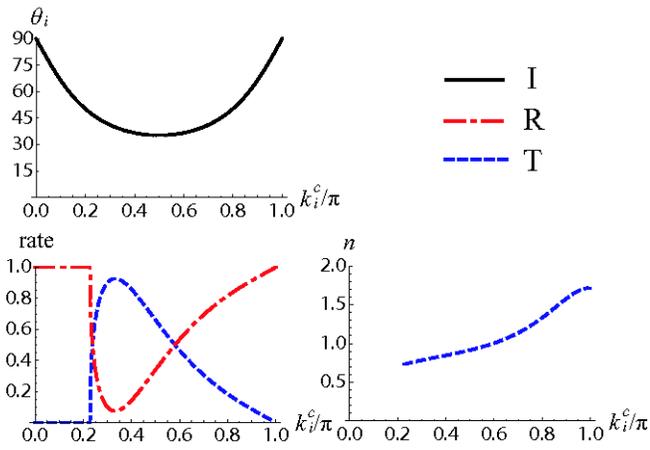}
\caption{(Color online) 
The incident angle $\theta_i$, the transmittance, the reflectance (rate), and the refraction index $n=\sin \theta_i/\sin\theta_t$ in the usual ferromagnet/ferromagnet junction system for $J = -1$, $J^\prime=-1.5$, $\Delta = -1.5$, $\mu_1/2S=-1.6$, $\mu_2/2S=-1.7$. 
The incident spinwave with $k_a = \pi/4$ and $k_b= 0$ in the usual ferromagnet is assumed. 
$I$, $R$, $T$ respectively denotes 
the incident wave, the reflection wave and the transmission wave.
The incident angle and the refraction index are determined by the direction 
of the group velocity. 
\label{Fig:usual_ref}}
\end{center}
\end{figure}

Let us study the reflectance and the transmittance of spinwaves
in the usual ferromagnet/ferromagnet junction system on the cubic lattice
with the boundary condition (\ref{SS:boundary}) (see Fig.~\ref{Fig:FMFM}). 
The Hamiltonian of the total junction system is given by
\begin{equation}
\begin{split}
&H_{\text{tot}1}=\sum_{\VEV{\boli,\bolj}\ \text{for}\ {i_c,j_c}\leq0}J{\bf S}_{\boli}\cdot {\bf S}_{\bolj}+\sum_{i_c\leq0}(-K_1(S_{\boli}^{z})^2+{\rm H}_1S_{\boli}^z)\\
&+\sum_{\VEV{\boli,\bolj}\ \text{for}\ {i_c,j_c}\geq 1}J^\prime{\bf S}_{\boli}\cdot {\bf S}_{\bolj}+\sum_{i_c\geq 1}(-K_2(S_{\boli}^{z})^2
+{\rm H}_2S_{\boli}^z)\\
&+\Delta\sum_{a,b}\bolS_{\boli=(a,b,c=0)} \cdot \bolS_{\bolj=(a,b,c=1)}\ ,
\end{split}
\end{equation}
where we assume that the boundary is located between $c=0$ and $1$, 
and $\VEV{i,j}$ represents the pairs of the nearest neighbor coupling. 
The reflection and refraction process with the incident spinwave of $\bolk_i$
may be written as the ket 
\begin{equation}
\begin{split}
\ket{RR_1}=&\sum_{a,b}e^{i(k_i^a a+k_i^b b)}
(A\sum_{c=-\infty}^{0}e^{ik_i^c c}\alpha^\dagger_{a,b,c}
+B\sum_{c=-\infty}^{0}e^{-ik_r^c c}\alpha^\dagger_{a,b,c}\\
&+C\sum_{c=1}^{\infty}e^{ik_t^c (c-1)}\alpha^\dagger_{a,b,c})\ket{0}\ ,
\end{split}
\end{equation}
where $k_t^c$ is given by eq.~(\ref{nom_kt}).
Then, $A$, $B$, $C$ are determined by the Schr\"{o}edinger equation:
\begin{equation}
H_{\text{tot}1}\ket{RR_1}=\Omega\ket{RR_1}\ .
\label{RReigen}
\end{equation}
If we consider the transition matrix element of $H_{\text{tot}1}\ket{RR_1}$ to 
a lattice position, (\ref{RReigen}) is always satisfied except at $c=0,1$.
Hence, we consider 
\begin{equation}
\begin{split}
\bra{0}a_{a,b,c=0} H_{\text{tot}1}\ket{RR_1}&=\Omega (A+B)e^{i(k_i^a a+k_i^b b)}\ ,\\
\bra{0}a_{a,b,c=1} H_{\text{tot}1}\ket{RR_1}&=\Omega Ce^{i(k_i^a a+k_i^b b)}\ .
\end{split}
\end{equation} 
Equivalently, 
\begin{equation}
\begin{split}
&(\xi_{1}(k_i^c)-J e^{-ik^c_i})A+(\xi_{1}(k_i^c)-J e^{ik^c_i})B=
\Delta C\ ,\\
&\Delta(A+B)=(\xi_{2}(k_t^c)-J^\prime e^{ik^c_t})C\ ,
\end{split}
\end{equation}
where 
\begin{equation}
\begin{split}
\xi_{1}(k_i^c)&= 2J(\cos k_i^c-1)-J+\Delta\ ,\\
\xi_{2}(k_t^c)&= 2J^\prime(\cos k_t^c-1)-J^\prime+\Delta\ .
\end{split}
\end{equation}
Hence, we obtain
\begin{equation}
\begin{split}
\frac{B}{A}&=-\frac{(\xi_{1}(k_i^c)-J e^{-ik^c_i})(\xi_{2}(k_t^c)-J^\prime e^{ik^c_t})-\Delta^2}{(\xi_{1}(k_i^c)-J e^{ik^c_i})(\xi_{2}(k_t^c)-J^\prime e^{ik^c_t})-\Delta^2}\ , 
\\
\frac{C}{A}&=-\Delta\frac{2iJ\sin k_i^c }{(\xi_{1}(k_i^c)-J e^{ik^c_i})(\xi_{2}(k_t^c)-J^\prime e^{ik^c_t})-\Delta^2}\ .
\label{BandC}
\end{split}
\end{equation}
By using the relation
\begin{equation}
\begin{split}
&|(\xi_{1}(k_i^c)-J e^{-ik^c_i})(\xi_{2}(k_t^c)-J^\prime e^{ik^c_t})-\Delta^2|^2\\
&-|(\xi_{1}(k_i^c)-J e^{ik^c_i})(\xi_{2}(k_t^c)-J^\prime e^{ik^c_t})-\Delta^2|^2\\
&=
4JJ^\prime\Delta^2\sin k_i^c\sin k_t^c=\Delta^2 
v_g^{(s1)}(\bolk_i)v_g^{(s2)}(\bolk_t)/S^2\ ,
\end{split}
\end{equation}
the magnon-number-conservation law given by eq.~(\ref{conserv_s}) is easily 
confirmed as
\begin{equation}
v_g^{(s1)c}(\bolk_i)(1-|\frac{B}{A}|^2)=v_g^{(s2)c}(\bolk_t)|\frac{C}{A}|^2\ ,
\end{equation}
regardless of any $\bolk_i$ and $\bolk_t$.
From eq.~(\ref{BandC}), 
we explicitly find the dependence of the transmittance and the phase shift of spinwave on the boundary condition $\Delta$. 
For example, the transmittance 
$|v_g^{(s2)c}(\bolk_t)/v_g^{(s1)c}(\bolk_i)||C/A|^2$, the reflectance 
$|B/A|^2$ and the refraction index $\sin\theta_i/\sin\theta_t$ 
in the specific case 
are shown in Fig.~\ref{Fig:usual_ref}.  
Since the travelling direction of spinwave
is given by the group velocity,
incident spinwave of different frequencies 
could have the same incident angle.
For $\Delta\rightarrow 0$, the transmittance leads: 
$|C/A|^2=O(\Delta^2/J^{\prime2})$.

\subsection{$J_1$-$J_2$ chains parallel to the boundary}
\begin{figure}[ht]
\begin{center}
\includegraphics[scale=0.5]{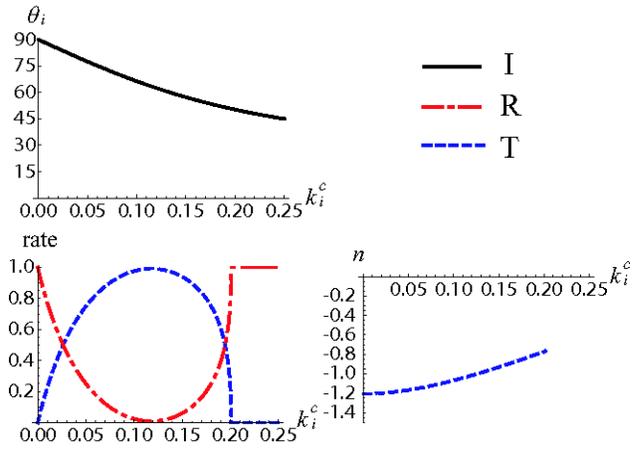}
\caption{(Color online) 
The incident angle $\theta_i$, the transmittance, the reflectance (rate), 
and the refraction index $n$ in the usual ferromagnet/ frustrated ferromagnet junction system 
consisting of the $J_1$-$J_2$ chains parallel to the boundary plane for
$J = -1$, $J_1=-2$, $J_2 = 1$, $J_3=-0.5$, 
$\Delta = -1.0$, $\mu_1/2S=-1.6$, $\mu_2/2S=-1.5$. 
The incident spinwave with $k_a = \pi/4$ and $k_b= 0$ in the usual ferromagnet is assumed. 
$I$, $R$, $T$ respectively denotes 
the incident wave, the reflection wave and the transmission wave.
The incident angle and the refraction index are determined by the direction 
of the group velocity. 
\label{Fig:parallel_ref}}
\end{center}
\end{figure}

Next, let us study the case that one side of 
the ferromagnets-junction system consists of the $J_1$-$J_2$ chains 
parallel to the boundary as shown in Fig.~\ref{Fig:FMJ1J2a}. 
The reflection and refraction process is described
by the following ket:
\begin{equation}
\begin{split}
\ket{RR_2}&=\sum_{a,b}e^{i(k_i^a a+k_i^b b)}
(A\sum_{c=-\infty}^{0}e^{ik_i^c c}\alpha^\dagger_{a,b,c}
+B\sum_{c=-\infty}^{0}e^{-ik_r^c c}\alpha^\dagger_{a,b,c}\\
&+C\sum_{c=1}^{\infty}e^{ik_t^c (c-1)}\alpha^\dagger_{a,b,c})\ket{0}\ ,
\end{split}
\end{equation}
where $k_t^c$ is given by eq.~(\ref{ktcJ1J21}). 
The procedure to determine $B,C$ is exactly the same as 
that of the previous section. 
Hence, $B$ and $C$ are given by (\ref{BandC}). 
For exapmle, the transmittance 
$|v_g^{(11)c}(\bolk_t)/v_g^{(s1)c}(\bolk_i)||C/A|^2$, the reflectance 
$|B/A|^2$ and the refraction index in the specific case 
are shown in Fig.~\ref{Fig:parallel_ref},  
where the occurrence of
 the negative refraction index is explicitly seen.

\subsection{$J_1$-$J_2$ chains perpendicular to the boundary}
\begin{figure}[ht]
\begin{center}
\includegraphics[scale=0.5]{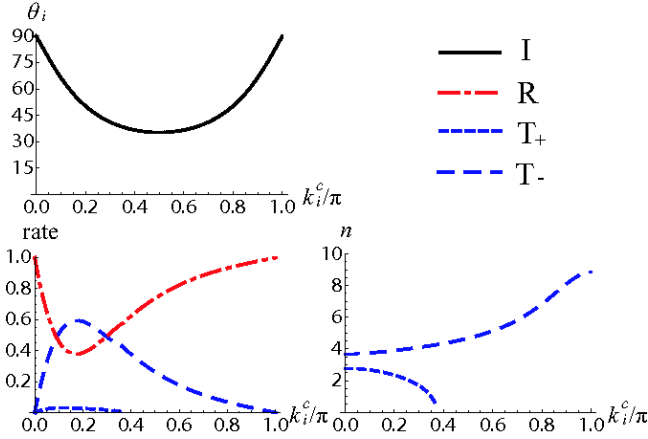}
\caption{(Color online) 
The incident angle $\theta_i$, the transmittance, the reflectance (rate), and the refraction index $n$ in the usual ferromagnet/ frustrated ferromagnet junction system 
consisting of the $J_1$-$J_2$ chains perpendicular to the boundary plane for
$J = -1$, $J_1=-1.5$, $J_2 = 1$, $J_3=-0.5$, 
$\Delta = -1.0$, $\mu_1/2S=-1.6$, $\mu_2/2S=-1.5$. 
The incident spinwave with $k_a = \pi/4$ and $k_b= 0$ in the usual ferromagnet is assumed. 
$I$, $R$, $T_{\pm}$ respectively denotes 
the incident wave, the reflection wave and the transmission wave with the wavevector $k^{(\pm)c}_t$ in eq.~(\ref{ktcJ1J22}).
The incident angle and the refraction index are determined by the direction 
of the group velocity.
\label{Fig:perpendicular_ref}}
\end{center}
\end{figure}

Finally, we consider the junction system of the $J_1$-$J_2$ chains 
perpendicular to the boundary as shown in Fig.~\ref{Fig:FMJ1J2b}. 
In this case, the ket is given by
\begin{equation}
\begin{split}
&\ket{RR_3}=\sum_{a,b}e^{i(k_i^a a+k_i^b b)}
(A\sum_{c=-\infty}^{0}e^{ik_i^c c}\alpha^\dagger_{a,b,c}
+B\sum_{c=-\infty}^{0}e^{-ik_r^c c}\alpha^\dagger_{a,b,c}\\
&+C_1\sum_{c=1}^{\infty}e^{ik_{t}^{(+)c} (c-1)}\alpha^\dagger_{a,b,c}
+C_2\sum_{c=1}^{\infty}e^{ik_{t}^{(-)c} (c-1)}\alpha^\dagger_{a,b,c})\ket{0}\ ,
\end{split}
\end{equation}
where $k_{t}^{(\pm)c}$ are given by eq.~(\ref{ktcJ1J22}). 
We see that (\ref{RReigen}) is satisfied at any lattice cite except at $c=0,1,2$.
Hence, we consider
\begin{equation}
\begin{split}
\bra{0}a_{a,b,c=0}H_{\text{tot}3}\ket{RR_1}&=\Omega(A+B)e^{i(k_i^a a+k_i^b b)}\ ,\ \ \\
\bra{0}a_{a,b,c=1}H_{\text{tot}3}\ket{RR_1}&=\Omega(C_1+C_2) e^{i(k_i^a a+k_i^b b)}\ ,\\
\bra{0}a_{a,b,c=2} 
H_{\text{tot}3}\ket{RR_1}&=\Omega(C_1e^{ik_{t}^{(+)c}}+C_2e^{ik_{t}^{(-)c}}) e^{i(k_i^a a+k_i^b b)}\ ,
\end{split}
\end{equation} 
where $H_{\text{tot}3}$ is the Hamiltonian of the total system considered in 
this subsection.  
Equivalently, 
\begin{equation}
\begin{split}
&(\xi^\prime_1(k_i^c)-Je^{-ik^c_i})A+(\xi^\prime_1(k_i^c)-Je^{ik^c_i})B=
\Delta (C_1+C_2)\ ,\\
&(\xi^\prime_2(k^{(+)c}_{t})-J_1e^{ik^{(+)c}_{t}}-J_2e^{i2k^{(+)c}_{t}})C_1\\
&+(\xi^\prime_2(k^{(+)c}_{t})-J_1 e^{ik^{(-)c}_{t}}-J_2e^{i2k^{(-)c}_{t}})C_2
=\Delta(A+B)\ ,\\
&(\xi^\prime_3(k^{(+)c}_{t})-2J_1\cos k^{(+)c}_{t}-J_2 e^{i2k^{(+)c}_{t}} )C_1e^{ik_{t}^{(+)c}}\\
&+(\xi^\prime_3(k^{(+)c}_{t})-2J_1\cos k^{(-)c}_{t} -J_2 e^{i2k^{(-)c}_{t}})C_2e^{ik_{t}^{(-)c}}=0\ ,
\end{split}
\end{equation}
where
\begin{equation}
\begin{split}
\xi^\prime_1(k^{c}_{i})&=2J(\cos k_i^c-1)-J+\Delta\ ,\\
\xi^\prime_2(k^{(+)c}_{t})&=\xi^\prime_2(k^{(-)c}_{t})=2(J_1\cos k^{(+)c}_{t}+J_2\cos 2k^{(+)c}_{t}
+\frac{J_1^2}{8J_2})\\
&\hspace{2cm}-J_1+J_2+\Delta\ ,\\
\xi^\prime_3(k^{(+)c}_{t})&=\xi^\prime_2(k^{(+)c}_{t})+J_1-\Delta\ .
\end{split}
\end{equation}
$B/A,C_1/A,C_2/A$ are given by solving these equations. 
We numerically confirmed the following magnon number conservation law 
in the various concrete parameters: 
\begin{equation}
v_g^{(s1)c}(\bolk_i)(1-|\frac{B}{A}|^2)=v_g^{(12)c}(\bolk_t^{(+)})|\frac{C_1}{A}|^2
+v_g^{(12)c}(\bolk_t^{(-)})|\frac{C_2}{A}|^2\ ,
\end{equation}
where $v_g^{(12)c}(\bolk)=-2S(J_1 \sin k^c+J_2 \sin 2k^c) $. 
For example, the transmittance 
$|v_g^{(12)c}(\bolk_t^{+})/v_g^{(s1)c}(\bolk_i)||C_1/A|^2$, 
$|v_g^{(12)c}(\bolk_t^{-})/v_g^{(s1)c}(\bolk_i)||C_2/A|^2$, 
the reflectance 
$|B/A|^2$ and the refraction index in the specific case 
are shown in Fig.~\ref{Fig:perpendicular_ref},  
where two species of transmission wave 
appear when both $k^{(\pm)c}_t$s in eq.~(\ref{ktcJ1J22})
have real values.

\section{Conclusion}
We studied the reflection and refraction process 
of spinwaves in the ferromagnet/frustrated ferromagnet 
junction system 
whose ferromagnets are described 
by the spin Hamiltonian with the generic Heisenberg-exchange coupling, 
the uniaxial anisotropic interaction and the external magnetic field. 
By using the Holstein-Primakoff spinwave expansion and 
taking the large $S$ limit, 
the ferromagnetic phase is given by the free bosonic Hamiltonian, 
which describes the dynamics of magnons. 
If frustration exists, the dispersion relation of magnons (spinwave) 
can become nontrivial. 

In Sec.~\ref{Sec:Snell}, we discussed the Snell's law of the spinwave 
in the case that the following conditions are 
satisfied in the junction system:
(i) The exchange interactions in each ferromagnet and on the boundary have a rotational symmetry around the $z$ direction in spin space. 
(ii) Each ferromagnet has a translational symmetry (far away from the boundary).
(iii) Both ferromagnets have the same lattice vectors in the $a$-$b$ plane, 
and 
the translational symmetry in the $a$-$b$ plane exists in the whole junction system. 
\cite{footnote}
By studying the various junction systems satisfying these conditions, 
we found the nontrivial refraction process, 
e.g., the splitting of the spinwave and the negative refraction. 

In Sec.~\ref{reftra}, we studied the `reflectance', `transmittance' 
and `phase shift at the boundary' in the concrete examples of 
the junction systems. 
Within the large $S$ limit, we exactly obtain these 
quantities, 
which explicitly depend on the boundary condition. 
Throughout this paper 
our results does not need the long wavelength approximation.

\begin{acknowledgment}
We thank G. Tatara, S. Murakami and A.~Yamaguchi for useful discussions.

\end{acknowledgment}

\appendix

\section{Current of angular momentum}
\label{Sec:current}
In this appendix, we briefly review the correspondence between 
the group velocity of spinwave and the current of angular momentum. 
We study the spin system described by the Hamiltonian (\ref{HeisenbergH}). 
Now, the Hamiltonian (\ref{HeisenbergH}) commutes with the total 
spin $S^z_{\text {tot}}=\sum_l S^z_{\bf j}$. Hence, $S^z_{\text {tot}}$ is 
the conserved quantity. This implies that the total magnon number $\sum_\boll \alpha^\dagger_\boll \alpha_\boll$ conserves, and the conserved current of magnons (related to the U(1) symmetry) exists. 

\begin{figure}[ht]
\begin{center}
\includegraphics[scale=0.3]{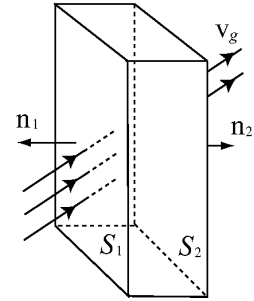}
\caption{Magnon current of the group velocity $v_g$ passes through the cuboid $C_1$. 
\label{Fig:magC}}
\end{center}
\end{figure}

The magnon current is understood by studying in- and outflow of the magnon number in a given region. 
Let us consider the cuboid $C_1$ of which two face-to-face surfaces $S_1$ and $S_2$ are infinitely large as shown in Fig.~\ref{Fig:magC}. 

The spatially-averaged flow of magnons is determined by the current passing through $S_1$ and $S_2$. 
By using the relation
\begin{equation}
[\alpha^\dagger_\boli \alpha_\boli+\alpha^\dagger_\bolj \alpha_\bolj, 
\bolS_\boli \cdot \bolS_\bolj]=0 \ ,
\end{equation}
we obtain 
\begin{equation}
i\frac{d(\sum_{l\in C_1} \alpha^\dagger_\boll \alpha_\boll)}{dt} = 
[\sum_{l\in C_1}\alpha^\dagger_\boll \alpha_\boll,H]=S\sum_{i\in C_1,\ j\in\hspace{-1.5mm} /C_1} J_{\boli\bolj}(\alpha^\dagger_\boli \alpha_\bolj -\alpha_\boli \alpha^\dagger_\bolj)\ ,
\end{equation}
where we use eq.~(\ref{Holst}). We assume that $J_{\boli\bolj}=J_\boll$ 
is short range so that $J_\boll=0$ for a large $|\boll|$, 
and that $S_1$ is sufficiently separated from $S_2$. 
Then, if we consider the state $\ket{\bolk}=\alpha^\dagger_\bolk \ket{0}$, 
we obtain 
\begin{equation}
\begin{split}
\bra{\bolk}\frac{d(\sum_{l\in C_1} \alpha^\dagger_\boll \alpha_\boll)}{dt}\ket{\bolk}
&=\frac{2S}{N}\sum_{\boli\in C_1,\ \bolj\in\hspace{-1.5mm} /C_1} 
J_{\boli\bolj}\sin \bolk\cdot(\boli-\bolj)\\
&=\frac{1}{N}(A_{S_1}\boln_1\cdot \bolv_g(\bolk)+A_{S_2}\boln_2\cdot \bolv_g(\bolk))\ ,
\end{split}
\end{equation}
where $A_{S_i}$ is the area of $S_i$ and 
\begin{equation}
\bolv_g(\bolk)\cdot\boln=2S\frac{\partial \epsilon(\bolk)}{\partial \bolk}\cdot \boln=-2S\sum_\boll J_\boll \sin(\bolk\cdot \boll) \boll\cdot\boln \ .
\end{equation}
These equations imply that
the group velocity can be viewed as the magnon current which carry the angular momentum. 
Even if there is a superposition of various spinwaves as $\sum_l A_l 
\alpha^\dagger_{\bolk_l}\ket{0}$, the time-averaged magnon current is simply summed as
\begin{equation}
\bar{\bolv}_g^{\text{tot}}=\sum_l |A_l|^2 \bolv_{g}(\bolk_l)\ .
\end{equation}


\begin{thebibliography}{9}
\bibitem{foot_0}
We use the terminology `trivial dispersion relation' 
when the dispersion relation is isotropic 
and the phase velocity is parall 
to the energy current (e.g., the Poynting vector in the case of light).
\bibitem{footnote}
{If we consider a junction of different compounds, the assumption (iii) 
is scarcely satisfied. However, if the difference of the lattice vectors in each compound is small, the deviation from the discussion in our paper may be perturbative and small. 
In addition, the translational invariance exists within the 
least common multiple lattice vectors in the $a$-$b$ plane. 
Hence, even if the difference of the lattice vectors is large, 
in some cases our simple discussion may predict the reasonable result, 
e.g., in the case of the long wavelength limit.}
\bibitem{footnote_boundary}
{As will be discussed, the refraction angles 
are independent of the concrete form of the boundary condition 
as long as the assumptions (i), (iii) hold: 
the Snell's law is independent of $\Delta$.
Only the transmission rate and the phase shift at the boundary depend on 
$\Delta$ as will be discussed in Sec.~\ref{ref_trans}.}
\end{thebibliography}
\end{document}